\documentclass[journal]{IEEEtran}
\usepackage{cite}

\usepackage{amssymb}
\usepackage{amsmath}
\usepackage{makeidx}
\usepackage{multicol}
\usepackage[abs]{overpic}
\usepackage{tabularx}
\usepackage{amsfonts}
\usepackage{array,color}
\usepackage{tabularx}
\usepackage{multirow}
\usepackage{graphics}
\usepackage{epsfig}
\usepackage{graphicx}
\usepackage{makeidx}
\usepackage{multicol}
\usepackage{amsthm}
\usepackage{bm}
\usepackage{booktabs}
\usepackage{ulem}
\usepackage{enumerate}
\usepackage[caption=false,font=footnotesize]{subfig}
\usepackage[font=small,labelsep=period]{caption}

\newtheorem{proposition}[theorem]{Proposition}

\usepackage{ulem}

\begin{document}

\title{Efficient Nonlinear Precoding for Massive MIMO Downlink Systems with 1-Bit DACs}
\author{Lei Chu, \IEEEmembership{Student Member, IEEE}, Fei Wen, \IEEEmembership{Member, IEEE}, Lily Li and  Robert Qiu, \IEEEmembership{Fellow, IEEE}
\IEEEcompsocitemizethanks{
\IEEEcompsocthanksitem Lei Chu is with Department of Electrical Engineering at Shanghai Jiaotong University, Shanghai, China. (leochu@sjtu.edu.cn)
\IEEEcompsocthanksitem Fei Wen is with the Department of Electronic Engineering, Shanghai Jiao Tong University, Shanghai 200240, China. (Corresponding author: Fei Wen)
\IEEEcompsocthanksitem Lily Li and Dr. Qiu are with the Department of Electrical and Computer Engineering\, Tennessee Technological University, Cookeville, TN 38505 USA. Dr. Qiu is also with Department of Electrical Engineering, Research Center for Big Data and Artificial Intelligence Engineering and Technologies, Shanghai Jiaotong University, Shanghai 200240, China. (e-mail: rqiu@ieee.org). \protect
}
\thanks{Dr. Qiu's  work  is partially supported  by  N.S.F. of China under Grant No.61571296 and N.S.F. of US under Grant No. CNS-1619250.
Dr. Wen's  work  is supported  by  N.S.F. of China under Grant No.61871265.}}

\IEEEtitleabstractindextext{%
\begin{abstract}
The power consumption of digital-to-analog converters (DACs) constitutes
a significant proportion of the total power consumption in a massive multiuser
multiple-input multiple-output (MU-MIMO) base station (BS).
Using 1-bit DACs can significantly reduce the power consumption.
This paper addresses the precoding problem for the massive narrow-band MU-MIMO downlink system
equipped with 1-bit DACs at each BS.
In such a system, the precoding problem plays a central role as the precoded symbols
are affected by extra distortion introduced by 1-bit DACs.
In this paper, we develop a highly-efficient nonlinear precoding
algorithm based on the alternative direction method framework.
Unlike the classic algorithms, such as the semidefinite relaxation (SDR)
and squared-infinity norm Douglas-Rachford splitting (SQUID) algorithms,
which solve convex relaxed versions of the original precoding problem,
the new algorithm solves the original nonconvex problem directly.
The new algorithm is guaranteed to globally converge under some mild conditions.
A sufficient condition for its convergence has been derived.
Experimental results in various conditions demonstrated that,
the new algorithm can achieve state-of-the-art performance comparable to the SDR algorithm,
while being much more efficient (e.g., more than 300 times faster than the SDR algorithm).

\end{abstract}

}

\maketitle

\IEEEpeerreviewmaketitle
\IEEEdisplaynontitleabstractindextext

\section{Introduction}
\label{Sec:a}
In recent years, there has been an increasing research attention
in MU-MIMO systems, in which a base station (BS) is equipped with a large number of antennas,
simultaneously serving many users terminals (UTs) in the same frequency band \cite{Liu2012Downlink}, \cite{Qiu2013Cognitive}.
Scaling up the number of the transmit antennas will lead to much more degrees of freedom available for each UT,
and result in higher data rate and improved radiated energy efficiency.
These benefits can be attained by simple linear processing such as maximum ratio transmission (MRT),
zero-forcing (ZF) or water filling (WF) on the downlink \cite{Rusek2012Scaling}, \cite{Joham2005Linear}.
Accordingly, massive MU-MIMO is foreseen as a promising technology for 5G cellular systems. 
In order to make full use of the favorable properties that massive MU-MIMO can offer,
one needs to pay attention to the related disadvantage caused by the deployment of
a large number of BS antennas.
For example, the total energy consumption grows linearly with the increase in the number of antennas,
and exponentially with the increase in the number of quantization bits
\cite{Usman2016MMSE, Swindlehurst2017Minimum}.
In this paper, we consider 1-bit digital-to-analog converters (DACs) for massive MU-MIMO systems,
which have the capability of dramatically reducing the hardware cost and energy consumption.

\subsection{Related Works}
\label{rw}

The 1-bit precoding schemes in current literature for downlink wireless communication systems are fundamentally based on
the well-known Bussgang theorem \cite{Julian1952Crosscorrelation},
with which one can decompose the nonlinear quantized signal output into
a linear function of the input to the quantizers and an uncorrelated distortion
term \cite{Mezghani2007On, Ucu2017Performance, Gokceoglu2017Spatio}.
With the novel decomposition and  using proper signal processing techniques, the quantized precoding problem is then tackled based on different performance metrics, e.g., sum rate \cite{Saxena2016On}, \cite{Jacobsson2017Massive},
coverage probability \cite{Usman2017Joint},
and (coded or uncoded) bit error rate (BER) \cite{Guerreiro2016Use}.
Assuming perfect channel state information (CSI), an expression of the downlink achievable rate
for matched-filter precoding has been derived in \cite{Saxena2017Analysis}. The performance of quantized transceivers in MU-MIMO downlink has been investigated in \cite{Jedda2017Massive, Xu2017On, Zhang2017app}. Moreover, it has been shown in a recent study \cite{Li2017Downlink} that,
compared with massive MIMO systems using ideal DACs, the performance (sum rate)
loss in 1-bit frequency-flat MU-MIMO systems can be compensated by disposing approximately
2.5 times more antennas at the BS. The capacity of the 1-bit millimeter wave {MIMO} system has been analyzed in \cite{Mo2015Capacity, Mo2015High}. It has been shown that, the Gaussian assumption of the quantization error does not hold at the high SNR regime.

Generally, for a MU-MIMO downlink system with 1-bit DACs, the application of 1-bit DACs could greatly reduce the power consumption and
significantly simplify the industrial design of the device in the BS. On the other hand, the coarsely quantized MU-MIMO systems, especially with higher-order modulation schemes, have to pay a high price for performance loss in terms of BER.

Recently, a convex relaxation based nonlinear precoder is proposed in
\cite{Jacobsson2016Nonlinear},  which enables 1-bit MU-MIMO system
to work well not only with the QPSK signaling but also with higher-order modulations.
Then several representative nonlinear precoders have been put forward based on different optimization algorithms.
For example, in a recent work \cite{Jedda2016Minimum}, a linear programming based precoding method has been proposed to mitigate the inter-user-interference and channel distortion in a 1-bit downlink MU-MISO system with QPSK symbols. Meanwhile, for 1-bit MU-MIMO, semidefinite relaxation (SDR) based precoding
has been brought up in \cite{Jacobsson2016Quantized}. The SDR precoder has the advantage of sound theoretical guarantees and shows robust performance in both small-scale and large-scale MU-MIMO systems.  Nevertheless, the high computational complexity of the SDR method
is the major obstacle to its application in massive MU-MIMO with a large number of antennas.
To address this limitation, an efficient squared-infinity norm Douglas-Rachford
splitting (SQUID) based precoder has been proposed in \cite{Jacobsson2016Quantized},
\cite{Casta2017POKEMON}; Besides, it has been shown that the SQUID precoder can achieve comparable performance with SDR precoder in large-scale MU-MIMO systems.  Further, two more efficient methods, namely C1PO and C2PO, have been developed in \cite{Casta20171}. However, they involve more hyper-parameters which need to be tuned empirically for different conditions, e.g., system sizes and modulation modes.

The goal of this work is to develop an efficient and robust precoder that can achieve state-of-the-art performance both in BER and computational complexity. The main contributions are as follows.

\subsection{Contributions}

First, we propose an efficient algorithm to solve the nonlinear precoding problem
based on the alternative direction method of multipliers (ADMM) framework.
The nonlinear precoding problem is generally difficult to solve due to the
nonconvex constraint, which enforces the elements of the solution vector
to have a uniform modulus.
Unlike the SDR precoding algorithm solving convex relaxed versions of the original nonconvex precoding problem,
we solve the nonconvex precoding problem directly via first reformulating it into an unconstrained form.
The SQUID algorithm is also based on the ADMM framework,
but it involves double loops in the iteration procedure,
as an inner loop is needed to solve the proximal operator for the squared $\ell_\infty$-norm.
In comparison, our algorithm has a single loop and thus is more efficient.

Second, a sufficient condition for the convergence of the new algorithm has been derived.
Since the nonlinear precoding problem is nonconvex and nonsmooth,
the convergence condition of the new algorithm is important for its
stable implementation in applications.
The results indicate that, the new algorithm is globally convergent
under a rather weak condition, e.g., provided that the penalty parameter is selected sufficiently large.

Finally, we have compared the new algorithm with some state-of-the-art
algorithms via numerical simulations in various conditions. The results demonstrated that the new algorithm can achieve state-of-the-art BER performance while maintaining the advantages of low complexity.

Matlab codes for reproducing the results in this work are available at https://github.com/Leo-Chu/OneBit\_MIMO .

\subsection{Paper Outline and Notations}

The remainder of this paper is structured as follows.
Section \ref{Sec:b} introduces the mathematical assumptions and outlines the system models.
The quantized precoding problem and the corresponding precoding algorithms are briefly presented in Section \ref{Sec:c}.
We then present the proposed nonlinear precoder in Section \ref{Sec:d}.
In Section \ref{Sec:e}, numerical studies are provided to evaluate the efficiency and effectiveness
of the proposed nonlinear precoding algorithm.
The conclusion of this paper is given in Section \ref{Sec:f}.
For the sake of simplicity, auxiliary technical derivations are deferred to the appendix.

\textit{Notations:} Throughout this paper, vectors and matrices are given in
lower and uppercase boldface letters, e.g., $\bf{x}$ and $\bf{X}$, respectively.
We use the symbol ${\left[ {\bf{X}} \right]_{kl}}$ to denote the element at the $k$th row and $l$th column.
The symbols ${ \rm{diag} \left( {\bf{X}} \right) }$, ${ \rm{tr} \left( {\bf{X}} \right) }$,
${ \mathbb{E} \left[ {\bf{X}} \right] }$, ${{\bf{X}} ^\mathrm{*}}$,
and ${{\bf{X}} ^\mathrm{T}}$ denote the diagonal operator, trace operator, expectation operator,
the conjugate, and the transpose of ${\bf{X}}$, respectively.
We use $\Re \left( {\bf{x}} \right)$, $\Im \left( {\bf{x}} \right)$, ${\left\| {\bf{x}} \right\|_1}$,
and ${\left\| {\bf{x}} \right\|_2}$ to represent the real part, the imaginary part, $\ell_1$-norm,
$\ell_2$-norm of vector ${\bf{x}}$. The notation $\bf{A} \succeq 0$ means $\bf{A}$ is nonnegative definite.
The subdifferential of the function $f$ is denoted by $\partial f\left(  \cdot  \right)$. We use ${\mathop{\rm sign}\nolimits} \left(  \cdot  \right)$ to denote the sign of a quantity
with ${\mathop{\rm sign}\nolimits} \left( 0 \right) = 0$. The symbol $\bf{I}$ indicates an identity matrix with proper size. The distance from a point ${\bf{x}} \in {\mathbb{R}^n}$ to a subset $S \in {\mathbb{R}^n}$ is denoted by ${\rm{dist}}({\bf{x}},S): = \inf \{ {\left\| {{\bf{y}} - {\bf{x}}} \right\|_2}:{\bf{y}} \in S\} $. We respectively write the minimal and maximal eigenvalues of a matrix as ${\rm{ei}}{{\rm{g}}_{\min }}( \cdot )$ and ${\rm{ei}}{{\rm{g}}_{\max }}( \cdot )$. The element-wise division operation is denoted by $\oslash$. The symbol ${\bf{1}}$ is a column vector with all elements being 1.



\section{System Model}
\label{Sec:b}

\subsection{Mathematical Assumptions}

In this work, we conduct analysis under assumption of perfect synchronization of the transmit antennas and perfect recovery of frame synchronization at the UTs \cite{Ng2014Robust, Joham2005Linear, Saxena2017Analysis, Gokceoglu2017Spatio}. The channel remains unchanged within the transmission time. It is assumed that the BS has perfect CSI. We will relax such an assumption and investigate the sensitivity of the proposed nonlinear precoding algorithm in the case of imperfect CSI in Section \ref{Sec:e}. Besides, to efficiently acquire CSI at each user in the massive MU-MIMO downlink \cite{Marzetta2010Noncooperative}, \cite{Nam2013Full}, we adopt an open-loop training under time-division duplex (TDD) operation\footnote{We prefer TDD operation to frequency-division duplex (FDD) operation on account of limited transmission resources in large-scale MIMO system. Specially, the number of downlink resources needed for pilots is proportionate to the number of BS antennas in FDD and the number of users in TDD \cite{Jose2011Pilot}, respectively. And the number of BS antennas is typically much bigger than the number of UT in a massive MU-MIMO system.}.

\subsection{MU-MIMO Downlink System with 1-bit DACs}

\begin{figure}[!t]
\centering
\includegraphics[width=0.49\textwidth]{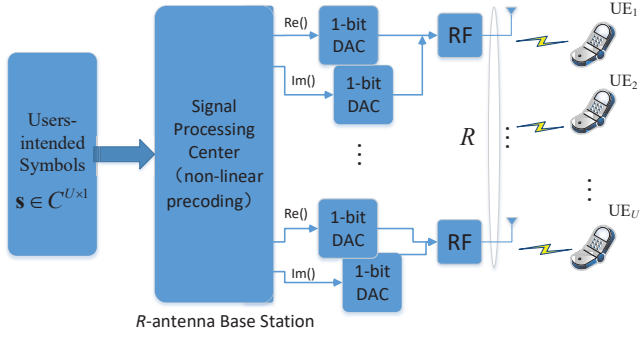}
\caption{The framework of the massive MU-MIMO downlink system with 1-bit DACs.}
\label{fig1}
\end{figure}

As shown in Fig. \ref{fig1}, we consider a single-cell quantized massive MU-MIMO downlink system with $U$ single-antenna UTs and a $R$-antenna base station (BS), in which each antenna is equipped with two 1-bit DACs. The signal processing center in Fig. \ref{fig1} controls the data processing procedure of the proposed 1-bit non-linear precoding scheme.

Let ${\bf{s}} \in {{\mathbb{C}}^{U \times 1}} $ be the constellation points to be sent to UTs. In our case (as shown Fig. \ref{fig1}), using the knowledge of CSI, the BS precodes ${\bf{s}}$ into a $R$-dimensional vector ${\bf{z}} = \mathcal{P}\left( {{\bf{H}},{\bf{s}}} \right)$, where $\mathcal{P}$ denotes an arbitrary, channel-dependent, mapping between the UT-intended symbols ${\bf{s}}$ and the precoded symbols ${\bf{z}}$. The precoded symbols satisfy the average power constraint \cite{Joham2005Linear}, \cite{Goldsmith:995349}
\begin{equation}
\label{eqi1}
{\mathbb{E}_{\bf{s}}}\left[ {{\bf{z}}^\mathrm{H}}{{\bf{z}}} \right] \le P_{TX}.
\end{equation}

The BS tries to send data symbols ${\bf{z}} \in {{\mathbb{C}}^{R \times 1}}$ simultaneously to UTs over $U \times R$ memoryless Gaussian block-fading channel $\bf{H}$. The input-output relationship of the MU-MIMO downlink system can be denoted as
\begin{equation}
\label{eq4}
{\bf{y}} =  {\bf{H}}{\bf{z}} + {\bf{n}},
\end{equation}
where the entries of ${\bf{ H}}$ are complex Gaussian random variables\footnote{Actually, the data analysis performed in the paper specifies no parameter distribution of the channel matrix, which implies a wide range of the practical applications.}, whose real and imaginary parts are assumed to be independent and identically distributed zero-mean Gaussian random variables with unit variance; ${{\bf{n}}}$ is a complex vector with element $n_i$ being complex addictive Gaussian noise distributed as ${{n_i}} \thicksim \mathcal{CN} \left( {0,{\varepsilon ^2} } \right)$ . The signal-to-noise ratio (SNR) is defined by $\alpha  = P_{TX}/{\varepsilon ^2}$.

The precoding problem based on ideal model with infinite resolution DACs has been investigated extensively by checking different performance metrics, i.e. throughput and mean square error (MSE) between received signal and transmitted signal. However, for the case of the quantized massive MU-MIMO \eqref{eq4}, the precoding problem should be revisited carefully as the precoded symbols at BS are affected by extra distortion introduced by low-resolution DACs. The detailed analysis is shown in the following.

\section{Quantized Precoding Problems and Solutions}
\label{Sec:c}

This section introduces the quantized precoding problem and briefly reviews representative precoding algorithms.

\subsection{Linear Quantized Problem and Solutions}

Let ${\bf{P}}$ be a precoding matrix and ${\bf x = Ps}$ the precoded symbols of an unquantized MU-MIMO sytem.
For the case of the massive MU-MIMO with 1-bit DACs, each precoded signal component $x_i, \ i = 1, \cdots, R$ is quantized separately into a finite set of prescribed labels by the symmetric uniform quantizer $Q$. It is assumed that the real and imaginary parts of precoded signals are quantized separately. The resulting quantized signals read
\begin{equation}
\label{eq3}
\begin{array}{c}
\begin{aligned}
{\bf{z}} &= Q({\bf{x}})\\
 &= Q(\Re \left( {\bf{x}} \right)) + jQ(\Im \left( {\bf{x}} \right))\\
 &= \kappa  \left( {\mathrm{sign}\left( {\Re \left( {\bf{x}} \right)} \right) + j \ \mathrm{sign}\left( {\Im \left( {\bf{x}} \right)} \right)} \right)
\end{aligned}
\end{array}.
\end{equation}
where $\kappa$ is set to $\kappa  = \sqrt {{P_{TX}}/2R} $ in order to satisfy the power constraint in \eqref{eqi1}. The output set of the 1-bit quantization is defined as $\Psi   = \kappa \left\{ {1 + j,1 - j, - 1 + j, - 1 - j} \right\}$.

For the linear precoding problem of the a MU-MIMO system as introduced in \eqref{eq4}, one can formulate the precoding problem by minimizing the MSE between the intended signal $\bf{s}$ and the quantized precoded signal $Q\left({\bf{Ps}}\right)$ through the channel $\bf{H}$ as follows \cite{Jacobsson2016Quantized}:
\begin{equation}
\label{eq5}
\begin{array}{*{20}{c}}
{\mathop {\rm minimize}\limits_{{\bf{P}} \in \mathbb{C}^{R \times U} , \ \rho } }& {\mathbb{E}_{\bf{s}}}[ {\left\| {{\bf{s}} - \rho {\bf{H}}Q\left( {{\bf{Ps}}} \right)} \right\|_2^2] + {\rho ^2}U{\varepsilon ^2}} \\
{{\mathop{\rm subject \ to}\nolimits} }&{\mathbb{E}_{\bf{s}}} [{\left\| {Q\left( {{\bf{Ps}}} \right)} \right\|_2^2] \le {P_{TX}},\rho  > 0}
\end{array}.
\end{equation}

Due to the nonlinearity, it is difficult to solve the problem \eqref{eq5} directly. Using the well-known Bussgang theorem \cite{Julian1952Crosscorrelation}, one can decompose the nonlinear signal output into a linear function of the input to the quantizers and an uncorrelated distortion term. Specially, we have
\begin{equation}
\label{eq6}
{\bf{z}} = Q\left( {{\bf{Ps}}} \right) = {\bf{GPs}} + {\bf{w}}.
\end{equation}

For the case of 1-bit DACs, under the assumption of
Gaussian input, i.e., ${\bf{s}} \sim \mathcal{CN}\left( {{\bf{0}},{{\bf{I}}}} \right)$, and the perfect CSI, it has been shown in the previous studies \cite{Usman2016MMSE, Jacobsson2016Quantized} that $\bf{G}$ can be expressed as:
\begin{equation}
\label{eq7}
{\bf{G}} = \sqrt {\frac{{2{P_{TX}}}}{{\pi R}}} {\mathop{\rm diag}\nolimits} {\left( {{\bf{P}}{{\bf{P}}^{\rm H}}} \right)^{ - 1/2}}.
\end{equation}
Substituting \eqref{eq6} and \eqref{eq7} into \eqref{eq5}, one can obtain the linear
quantized precoding matrix and the precoding factor by the
Lagrangian multiplier method, as shown in \cite{Joham2005Linear}.

\subsection{Nonlinear Quantized Problem and Solutions}
\label{NOPS1}

The use of 1-bit DACs could greatly reduce the power consumption and significantly simplify the industrial design of the device in the BS. However, the quantized MU-MIMO system, especially with higher-order modulation modes, suffers from heavy performance loss by applying linear 1-bit precoding algorithms. Besides, the BER curves achievable with linear precoding algorithms saturate at certain finite SNR, above which no further improvement can be obtained (See Section V-C for more details). To achieve
the goal of closing the performance gap\footnote{It is proven in recent study \cite{Li2017Downlink} that, compared with massive MIMO systems with ideal DACs, the sum rate loss in 1-bit massive MU-MIMO systems can be compensated for by disposing approximately 2.5 times more antennas at the BS. Accordingly, we will restrict us by developing possible precoding algorithms in regard to reducing bit error rate.}, nonlinear quantized
precoding algorithms have been developed.

\subsubsection{Nonlinear Quantized Precoding Problem}
\label{qpp}

In the 1-bit case, regardless of the precoding techniques, the outputs of DACs share the same amplitude:
\begin{equation}
\label{eq81}
{\left| {{{\left[ {\bf{z}} \right]}_{\rm{1}}}} \right|^2}{\rm{ = }} \cdots {\rm{ = }}{\left| {{{\left[ {\bf{z}} \right]}_R}} \right|^2} = {P_{TX}}/R.
\end{equation}
Under the minimum mean square error (MMSE) criterion, the 1-bit nonlinear precoding problem can be formulated as \cite{Jacobsson2016Quantized}
\begin{equation}
\label{eq8}
\begin{array}{*{20}{c}}
{\mathop {\rm minimize }\limits_{{\bf z}, \ \rho \in \mathbb{R}} }&{ \left\| {{\bf{s}} - \rho {\bf{Hz}}} \right\|_2^2 + {\rho ^2}U{\varepsilon ^2}}\\
{{\mathop{\rm subject \ to}\nolimits} }&{{\left| {{{\left[ {\bf{z}} \right]}_{\rm{1}}}} \right|^2}{\rm{ = }} \cdots {\rm{ = }}{\left| {{{\left[ {\bf{z}} \right]}_R}} \right|^2} = {P_{TX}}/R, \ \rho  > 0}
\end{array}.
\end{equation}

Representative nonlinear precoders for the precoding problem
(8) include the SDR precoder \cite{Jacobsson2016Quantized}, the SQUID precoder \cite{Jacobsson2016Quantized}, the C1PO/C2PO precoder\cite{Casta20171}, and the MSM precoder  \cite{Jedda2017Massive}, \cite{Jedda2016Minimum},
\cite{Jedda2018Quantized}), each with its own strengths and weaknesses. See Section \ref{rw} for more details. We highlight the SDR precoder as the benchmark nonlinear precoding algorithm as it has been shown in \cite{Sidiropoulos2006A, Kisialiou2010Proba} that SDR can provide a provably near-optimal solution, achieving a constant factor approximation of the optimal solution in probability. Besides, it has been shown in \cite{Jacobsson2016Quantized} that the SDR precoder has robust BER performance in both small-scale and large-scale MU-MIMO systems over a wide range of SNR. However, the disadvantage is that, with complexity of $O((2R+1)^{4.5})$ at the worst-case \cite{Boyd2006Convex}, \cite{Luo2010Semidefinite}, the SDR precoder is not suitable for high-dimensional problem, e.g., the massive MU-MIMO systems with hundreds of antennas. The discussion above motivates us to develop a precoding algorithm with high efficiency while being robust to the variation of system size and SNR.

\section{The Proposed Precoding Algorithm}
\label{Sec:d}

Unlike traditional methods solving relaxed formulations of \eqref{eq8} as it is nonconvex,
e.g., SDR [27] and $\ell_\infty$ relaxation \cite{Jacobsson2016Quantized},
in this section we present an efficient ADMM algorithm to directly solve the problem \eqref{eq8}
by first reformulating it into an unconstrained form.
Although the reformulated problem is nonconvex and nonsmooth\footnote{As the objective function
containing an indicator function on a nonconvex set is nonconvex and nonsmooth.},
we derive a sufficient condition under which the proposed ADMM algorithm is globally convergent.

\subsection{Algorithm}
\label{pa}
ADMM is a powerful framework well suited to solve many high-dimensional optimization problems
and has been widely applied in signal/image processing \cite{Wen2017Robust, Wen2018A, Wen2017Efficient}, statistics \cite{li2015global} and machine learning \cite{Hu2013Fast}. ADMM can be viewed as a special case of the
Douglas-Rachford method \cite{Douglas1956On} (equivalent to the Douglas-Rachford splitting method applied to a dual problem).
It utilizes a decomposition-coordination procedure to naturally decouple the variables in the global problem,
which makes the global problem easy to tackle.

First, denote ${\bf{v}} = {\bf{\rho z}}$ and
\begin{equation}
\label{eqd31}
\begin{array}{l}
\tilde{\bf{{s}}} = \left[ \begin{array}{l}
{\mathop{\rm \Re}\nolimits} \left( {\bf{s}} \right)\\
{\mathop{\rm \Im}\nolimits} \left( {\bf{s}} \right)
\end{array} \right],~~~
\tilde{\bf{{v}}} = \left[ \begin{array}{l}
{\mathop{\rm \Re}\nolimits} \left( {\bf{v}} \right)\\
{\mathop{\rm \Im}\nolimits} \left( {\bf{v}} \right)
\end{array} \right],\\
\tilde{\bf{{H}}} = \left[ {\begin{array}{*{20}{c}}
{{\mathop{\Re }\nolimits} \left( {\bf{H}} \right)}&{ - {\mathop{\Im}\nolimits} \left( {\bf{H}} \right)}\\
{{\mathop{\Im}\nolimits} \left( {\bf{H}} \right)}&{{\mathop{\Re}\nolimits} \left( {\bf{H}} \right)}
\end{array}} \right].
\end{array}
\end{equation}

Then, the complex-valued problem in \eqref{eq8} can be equivalently expressed as a real-valued problem as
\begin{equation}
\label{eqd4}
\begin{split}
&{\mathop {\mathrm{minimize} }\limits_{\tilde{\bf{{{v}}}}} }~~{\left\| {\tilde{\bf{{s}}} - \tilde{\bf{{H}}}  \tilde{\bf{{v}}}} \right\|_2^2 + \frac{{U{\varepsilon ^2}}}{P_{TX}} \left\| \tilde{\bf{{v}}} \right\|_2^2  }\\
&{\rm{subject~ to}}~~{\tilde{\bf{{v}}} \in {\Omega}}
\end{split},
\end{equation}
where $\Omega$ is a set defined as
\[ \Omega  = \left\{{\bf{x}} \in{\mathbb{R}^{2R}}: [ {\bf{{x}}} ]^2_1 =  \cdots  = [ {\bf{{x}}} ]^2_{2R} \right\} > 0. \]


Let $\delta_{\Omega} \left( \mathbf{o} \right)$ denote the indicator function defined on the set ${\Omega}$ as
\begin{equation}
\label{eqp1}
\delta_{\Omega} \left( \mathbf{o} \right) = \left\{ {\begin{array}{*{20}{c}}
{0,}&{ \ \mathbf{o} \ \in \ \Omega }\\
{\infty ,}&~~{ \mathrm{otherwise} }
\end{array}} \right. .
\end{equation}
Then, the problem \eqref{eqd4} can be rewritten as
\begin{equation} 
\label{eqp2}
{\mathop {\mathrm{minimize} }\limits_{\tilde{\bf{{{v}}}}} }~~ f\left(\tilde{\bf{{{v}}}}\right),
\end{equation}
where
\[f\left(\tilde{\bf{{{v}}}}\right) = {\left\| \tilde{\bf{{{s}}}} - \tilde{\bf{{{H}}}} \tilde{\bf{{{v}}}} \right\|_2^2 + c \left\| {\tilde{\bf{{{v}}}}} \right\|_2^2 + \delta_{\Omega} \left( {\tilde{\bf{{{v}}}}} \right)} ,\]
with $c=U\varepsilon^2/P_{TX}$.
With this new formulation \eqref{eqp2}, we can solve the problem \eqref{eq8}
using the efficient ADMM framework, which decouples the unknown variables and
tackles the problem in an efficient manner. Specially, the problem \eqref{eqp2}
can be equivalently expressed as
\begin{equation}
\label{eqp3}
\begin{split}
&{\mathop {\mathrm{minimize} }\limits_{\tilde{\bf{{{v}}}}} }~~{\left\| \tilde{\bf{{s}}} - \tilde{\bf{{H}}} {\tilde{\bf{v}}} \right\|_2^2 + c \left\| \tilde{\bf{{v}}} \right\|_2^2 + \delta_{\Omega} \left( {\bf{u}} \right)}\\
&{\rm{subject~ to}}~~ \tilde{\bf{{v}}} - {{\bf{u}} = {\bf{0}}}
\end{split},
\end{equation}
where ${\bf{u}}$ is an auxiliary variable.
The augmented Lagrangian associated with the problem \eqref{eqp3} is
\begin{equation}
\label{eqp4}
\begin{aligned}
{\mathcal{L}_\lambda }\left( {\tilde{\bf{{v}}},{\bf{u}}, {\bf{w}}} \right)  &= g\left(\tilde{\bf{{{v}}}}, {\bf{u}}\right)
 - \left<{{\bf{w}}}, {\tilde{\bf{{v}}}-{\bf{u}}} \right> + \frac{\lambda}{2} \left\| \tilde{\bf{{v}}}-{{\bf{u}}} \right\|_2^2\\
 &= g\left(\tilde{\bf{{{v}}}}, {\bf{u}}\right) + \frac{\lambda}{2}\left\| {\tilde{\bf{{v}}} -{\bf{u}} - \frac{\bf{w}}{\lambda}} \right\|_2^2 - \frac{\|{\bf{w}}\|^2_2}{2\lambda} ,
\end{aligned}
\end{equation}
where $\bf{w}$ is the dual variable, $\lambda>0$ accounts for the penalty factor related to the argumentation, and
\[g\left(\tilde{\bf{{{v}}}}, {\bf{u}}\right) =  \left\| {\tilde{\bf{{s}}} - \tilde{\bf{H}} {\tilde{\bf{v}}}} \right\|_2^2 + c \left\| \tilde{\bf{{v}}} \right\|_2^2 + \delta_{\Omega} \left( {\bf{u}} \right).\]

Then, ADMM applied to the problem \eqref{eqp4} consists of the following three steps in the $(k+1)$-th iteration
\begin{align}
\label{eq161}
{\tilde{\bf{{v}}}^{k + 1}} &= \arg\mathop { \min }\limits_{\tilde{\bf{{v}}}} \bigg( \left\| {\tilde{\bf{s}} - \tilde{\bf{{H}}} {\tilde{\bf{{v}}}}} \right\|_2^2 + c\left\| {{\tilde{\bf{{v}}}}} \right\|_2^2 \notag \\
&~~~~~~~~~~~~~~~~~~~~~~~ + \frac{\lambda}{2} \left\| {\tilde{\bf{{v}}}}-{{{\bf{u}}^{k}} - \frac{{\bf{w}}^k}{\lambda} } \right\|_2^2 \bigg), \\ \label{eq162}
{{\bf{u}}^{k + 1}} &= \arg\mathop { \min }\limits_{{\bf{u}}} \left( {\delta_{\Omega} \left( {{\bf{u}}} \right) + \frac{\lambda}{2} \left\| {{\tilde{\bf{{v}}}^{k+1}}- {{\bf{u}}} - \frac{{\bf{w}}^k}{\lambda} } \right\|_2^2} \right),
\\ \label{eq163}
{{\bf{w}}^{k + 1}} &= {{\bf{w}}^k} - \lambda\left( {\tilde{\bf{{v}}}^{k + 1}} - {{\bf{u}}^{k+1}}\right).
\end{align}

The $\tilde{\bf{{v}}}$-subproblem (15) is quadratic which has a closed-form solution given by
\begin{equation}
\label{eqp8}
{\tilde{\bf{{v}}}^{k+1}} = \left[2\tilde{\bf{{H}}}^{\rm{T}}\tilde{\bf{{H}}}+(2c + \lambda){\bf{I}} \right]^{-1}
\left(2\tilde{\bf{{H}}}^{\rm{T}}\tilde{\bf{{s}}}+\lambda{\bf{u}}^{k}+{\bf{w}}^k \right).
\end{equation}
Let  ${\boldsymbol{\omega }} = {{\bf{\tilde v}}^{k + 1}} - {{\bf{w}}^k}/\lambda $, ${\bf{u}}$-subproblem (16) is a projection of ${\boldsymbol{\omega }}$ onto the set $\Omega$, whose solution is given by (derived in Appendix A)
\begin{equation}
\label{eqp9}
{{\bf{u}}^{k + 1}} = \frac{{\left\| {\boldsymbol{\omega }} \right\|}}{{2R}}{\boldsymbol{\theta }},
\end{equation}
and
\begin{equation}
\label{eqq20}
{\theta _i} = \left\{ {\begin{array}{*{20}{l}}
{\{  - 1, + 1\} ,}&{{\omega _i} = 0}\\
{{\rm{sign(}}{\omega _i}{\rm{),}}}&{{\omega _i} \ne 0}
\end{array}} \right. .
\end{equation}
Note that, when some elements in ${\boldsymbol{\omega }}$  are zero, the solution is not unique. When all the elements in ${\boldsymbol{\omega }}$  are nonzero, the solution\footnote{In implementation we simply use \eqref{eqqq} since in practice we have never seen the occurrence of any elements in  ${\boldsymbol{\omega }}$ be zero.} reduces to
\begin{equation}
\label{eqqq}
{{\bf{u}}^{k + 1}} = {\rm{sign}}({\boldsymbol{\omega }})\frac{{\left\| {\boldsymbol{\omega }} \right\|_1}}{{2R}}.
\end{equation}

Finally, let ${\hat {\bf{u}}}$ denote the solution of the above ADMM algorithm, using the rounding strategy as described in \cite{Luo2010Semidefinite},
we can obtain the desired real-valued precoded vector by
\begin{equation}
\label{eqp10}
\hat{\bf{ z}} = \kappa \ {\mathop{\rm sign}\nolimits} \left[ {\hat{\bf{u}}} \right].
\end{equation}
Then, the complex-valued precoded vector can be determined as
\begin{equation}
\label{eqp11}
{\tilde{\bf{ z}}} = {\left[ {\hat{\bf{ z}}} \right]_{1:R}} + j {\left[ {\hat{\bf{ z}}} \right]_{R + 1:2R}}.
\end{equation}

In this ADMM algorithm, if the penalty parameter $\lambda$
does not change in the iteration procedure,
we only need to compute the matrix inversion in the $\tilde{\bf{{v}}}$-update \eqref{eqp8} once.
Besides, the dominant computational load in each iteration
is matrix-vector multiplication with complexity $O(R^2)$.

\subsection{Convergence Analysis}
\label{Sec:IV2}

As the set $\Omega$ is nonconvex, the problem \eqref{eqp2} is nonconvex and nonsmooth.
While the convergence properties of ADMM have been well established for the convex case,
there have been only a few studies reported very recently for the nonconvex case.
For the convex case, the convergence of the ADMM algorithm is easily guaranteed.
But for the nonconvex case, the convergence of the ADMM algorithm relies on the
property of the objective function and the selection of the penalty parameter.
In the following, we derive a sufficient condition for the convergence of the
proposed ADMM algorithm using the approach in \cite{li2015global}.

We first give three lemmas for deriving the sufficient condition.
The proofs are given in Appendix. In the sequel for convenience we use the notations:
${\varphi} = {\rm{ei}}{{\rm{g}}_{\max }}({{\tilde{\bf{H}}}^{\rm T}}{\tilde{\bf{H}}})$,
which denotes the maximal eigenvalues of ${{\tilde{\bf{H}}}^{\rm T}}{\tilde{\bf{H}}}$.

\textit{Lemma I}. The sequence $\{ ({\tilde{\bf{ v}}^k},{{\bf{u}}^k},{{\bf{w}}^k})\}$ generated via \eqref{eq161}--\eqref{eq163} satisfies
\begin{equation}
\begin{split}
&\mathcal{L}({{\tilde{\bf{ v}}}^{k + 1}},{{\bf{u}}^{k + 1}},{{\bf{w}}^{k + 1}}) \le \mathcal{L}({{\tilde{\bf{ v}}}^k},{{\bf{u}}^k},{{\bf{w}}^k})\\
&~~~~~~~~- \left[ {\frac{{ 2c + \lambda }}{2} - \frac{{4{{({\varphi} + c)}^2}}}{\lambda }} \right]\left\| {{{\tilde{\bf{ v}}}^{k + 1}} - {{\tilde{\bf{ v}}}^k}} \right\|_2^2. \notag
\end{split}
\end{equation}

In particular, if the condition
\begin{equation}\label{equ:conv_cond}
\lambda  > \max \left( {\sqrt {{{c}^2} + 8{{({\varphi} + c)}^2}}  - c,~8{\varphi},~8c} \right)
\end{equation}
 holds, $\mathcal{L}$ is monotonously decreasing in the iteration procedure.

Lemma I implies that, under the condition (\ref{equ:conv_cond}),
$\mathcal{L}$ is nonincreasing in the iteration and thus is convergent as it is lower semi-continuous.

\textit{Lemma II}. For the sequence $\{ ({\tilde{\bf{ v}}^k},{{\bf{u}}^k},{{\bf{w}}^k})\}$ generated via \eqref{eq161}--\eqref{eq163},
denote ${{\bf{z}}^k}: = ({\tilde{\bf{ v}}^k},{{\bf{u}}^k},{{\bf{w}}^k})$, suppose that (\ref{equ:conv_cond}) holds, then
\[\mathop {\lim }\limits_{k \to \infty } \left\| {{{\bf{z}}^{k + 1}} - {{\bf{z}}^k}} \right\|_2^2 = 0.\]
In particular, any cluster point of $\{ ({\tilde{\bf{ v}}^k},{{\bf{u}}^k},{{\bf{w}}^k})\}$ is a stationary point of $\mathcal{L}$.

\textit{Lemma III}. For the sequence $\{ ({\tilde{\bf{ v}}^k},{{\bf{u}}^k},{{\bf{w}}^k})\}$
generated via \eqref{eq161}--\eqref{eq163}, suppose that (\ref{equ:conv_cond}) holds, then there exists a constant $c_0>0$ such that
\[{\rm{dist}}(0,\partial \mathcal{L}({\tilde{\bf{ v}}^k},{{\bf{u}}^k},{{\bf{w}}^k})) \le {c_0}\left\| {{{\tilde{\bf{ v}}}^{k + 1}} - {{\tilde{\bf{ v}}}^k}} \right\|_2^2.\]

Lemma III establishes a subgradient lower bound for the iterate gap,
which together with Lemma II implies that
${\rm{dist}}(0,\partial \mathcal{L}({\tilde{\bf{ v}}^k},{{\bf{u}}^k},{{\bf{w}}^k})) \to 0$ as $k \to \infty$.

Based on the results in the above lemmas, we can conclude the following theorem.

\textit{Theorem I}. Suppose that the penalty parameter $\lambda$ satisfies the condition  \eqref{equ:conv_cond}, then the sequence $\{ ({\tilde{\bf{ v}}^k},{{\bf{u}}^k},{{\bf{w}}^k})\}$ generated by
the ADMM steps \eqref{eq161}--\eqref{eq163} converges to a stationary point of the problem \eqref{eqp2}.

\subsection{Efficient Implementation of the Proposed Algorithm}

This result in Theorem I implies that, the proposed ADMM algorithm is globally convergent
if the penalty parameter $\lambda$ is selected sufficiently large to satisfy (\ref{equ:conv_cond}).
To further accelerate the ADMM algorithm,
a standard trick is to use a continuation process for the penalty parameter $\lambda$. 
More specifically, we use a small starting value of $\lambda$ and
gradually increase it in the iteration procedure until reaching a target value
satisfying the condition \eqref{equ:conv_cond}, e.g., ${\lambda _0} \le  \cdots  \le {\lambda _K} = {\lambda _{K + 1}} =  \cdots  = \lambda$.
Theorem I still applies if the value of $\lambda$ becomes fixed after a finite number of iterations.

Besides, it is noted that the dynamic adjusting of $\lambda$ will
result in performing matrix inversion in \eqref{eqp8} in each iteration.
In fact, we only need to compute matrix inverse only once in the first iteration
by using the fact that ${\tilde{\bf{H}}}^{\rm{T}}{\tilde{\bf{H}}}$ is a symmetric matrix.
Specifically, in the first iteration, we can perform the singular value decomposition
(SVD) of ${\tilde{\bf{H}}}^{\rm{T}}{\tilde{\bf{H}}}$ to yield
$\left( {{\bf{U}},{\bf{S}},{{\bf{U}}^{\rm{T}}}} \right){\rm{ = svd}}\left( {\tilde{\bf{H}}}^{\rm{T}}{\tilde{\bf{H}}} \right)$;
${\bf{U}}$ contains the singular vectors and ${\bf{U}}{{\bf{U}}^{\rm{T}}} = {\bf{I}}$. 
Then, denote
${\left( {{\bf{S}} + \alpha {\bf{I}}} \right)^{ - 1}} = {\mathop{\rm diag}\nolimits} \left( {\bf{a}} \right)$,
using the matrix inversion formulas
\[{\left( {{\bf{ABC}}} \right)^{ - 1}} = {{\bf{C}}^{ - 1}}{{\bf{B}}^{ - 1}}{{\bf{A}}^{ - 1}},\]
\[{\left( {{\bf{US}}{{\bf{U}}^{\rm{T}}} + \alpha {\bf{I}}} \right)^{ - 1}} = {\bf{U}}{\left( {{\bf{S}} + \alpha {\bf{I}}} \right)^{ - 1}}{{\bf{U}}^{\rm{T}}},\]
and the relation
\[{\mathop{\rm diag}\nolimits} {\left( {\bf{a}} \right)^{ - 1}}{{\bf{U}}^{\rm T}\bf{c}} = {{\bf{U}}^{\rm T} \bf{c}} \oslash {\bf{a}},\]
we can efficiently update the ${\tilde{\bf{{v}}}}$-subproblem in the subsequent iterations as
\begin{equation}
\begin{split}
{\tilde{\bf{{v}}}^{k+1}} &= {\left[ {2{\tilde{\bf{H}}}^{\rm{T}}{\tilde{\bf{H}}} + (2c + {\lambda ^{k+1}}){\bf{I}}} \right]^{ - 1}} {\bf{d}}\\
 &= {\bf{U}}\left[ {\left( {{{\bf{U}}^{\rm{T}}} {\bf{d}}} \right) \oslash \hat{\bf{s}}} \right],
\end{split}
\end{equation}
where
\[{\bf{d}} = \left( {2{{\tilde{\bf{H}}}^{\rm{T}}}\tilde {\bf{s}} + {\lambda ^{k+1}}{{\bf{u}}^{k + 1}} + {{\bf{w}}^k}} \right), \]
and
\[\hat {\bf{s}} = 2{\mathop{\rm diag}\nolimits} \left( {\bf{S}} \right) + \left( {2c + {\lambda ^{k+1}}} \right){\bf{1}}.\]
In this manner, the dominant computation load in each subsequent iteration
is matrix-vector multiplication with computational complexity $O(R^2)$,
which improves the efficiency of the proposed algorithm.

\section{Simulations}
\label{Sec:e}

This section evaluates the performance of the new algorithm via numerical simulations
in comparison with six well-established precoders (C1PO\cite{Casta20171}, C2PO\cite{Casta20171}, SDR \cite{Jacobsson2016Quantized}, SQUID \cite{Casta20171}, \cite{Casta2017POKEMON}, MSM \cite{Jedda2016Minimum}, \cite{Jedda2017Massive}, \cite{ Jedda2016Minimum, Jedda2017Massive, Jedda2018Quantized}),
and one linear quantized precoder (ZF).
The proposed algorithm is initialized with zero,
and the penalty parameter $\lambda$ is selected to satisfy
the convergence condition \eqref{equ:conv_cond}.
Each provided result is an average over 1000 independent
runs (except for the results in Fig. 2, where only one simulation
for each condition to illustrate the convergence behavior).  The ZF precoder with infinite-resolution DACs (ZFi) is
regarded as the benchmark.

\subsection{Convergence Behavior of the Proposed Precoder}

First, we show the convergence behavior of the proposed algorithm
by investigating the iterate gap of the $\tilde{\bf{{v}}}$ and ${\bf{{u}}}$ variables,
defined as
\begin{equation}
\begin{split}
\Delta {\tilde{\bf{{v}}}^{k}} &=  \left\| {{\tilde{\bf{{v}}}^{k + 1}} - {\tilde{\bf{{v}}}^k}} \right\|_2 / \left\| {{\tilde{\bf{{v}}}^{k + 1}} } \right\|_2, \\
\Delta {{\bf{u}}^{k}} &=  \left\| {{{\bf{u}}^{k + 1}} - {{\bf{u}}^k}} \right\|_2 / \left\| {{{\bf{{u}}}^{k + 1}} } \right\|_2.
\end{split}
\end{equation}

\begin{figure}[!t]
\centering
\subfloat[Iterate gap of the ${\tilde{\bf{{v}}}}$ variable]{ \label{fig4a}
\includegraphics[width=0.49\textwidth]{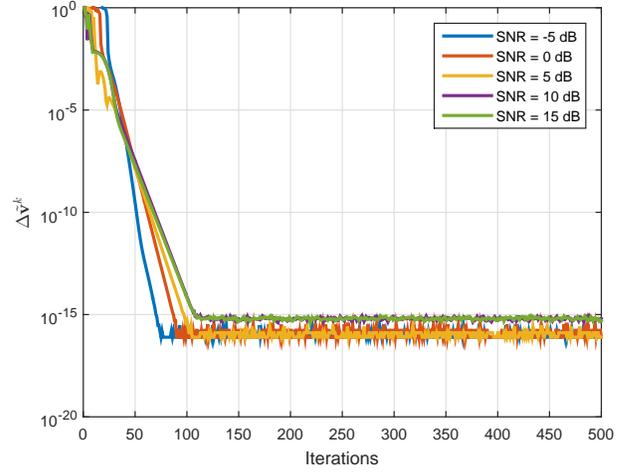}
}\\
\subfloat[Iterate gap of the ${{\bf{{u}}}}$ variable]{ \label{fig4b}
\includegraphics[width=0.49\textwidth]{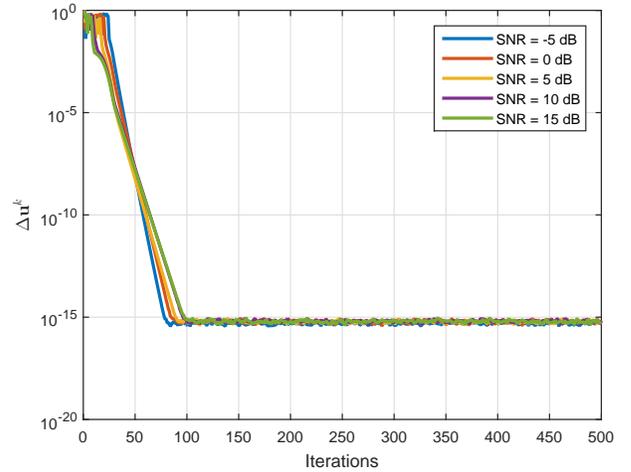}
}
\caption{Typical convergence behavior of the proposed algorithm in different SNR.}
\label{fig4x}
\end{figure}

We consider a MU-MIMO system with 16 BS antennas and 4 UTs,
with a wide range of the SNR values. Fig. \ref{fig4x}
shows the typical convergence behavior of the proposed precoder.
The results in Fig. \ref{fig4x} demonstrate that
the proposed algorithm is convergent and it can converge to
a sufficient accuracy within only tens of iterations.
For example, it can converge to an accuracy with tolerance lower than $10^{-7}$
within only 50 iterations in a wide range of SNR values.  Moreover, the plotted iteration gaps in Fig. \ref{fig4x} indicates that the new algorithm has an eventually linear convergence rate (namely, local linear convergence rate \cite{Lewis2009Local}).

\subsection{Computational Complexity Comparison}

We consider a $R$-antenna MU-MIMO system using 16QAM modulation. The number of UTs is $U = 10$. The numbers of antennas are
16, 32, 64, 128 and 256, respectively. The SDR based precoder has a computational
complexity of $O((2R+1)^{4.5})$ at the worst-case.
The SQUID precoder with two loops has a computational complexity of
$O(2(k_1 R^{3} + k_2 R^{2}))$ ($k_1$ and $k_2$ are the iteration numbers of the two loops). As shown in \cite{Casta20171}, computing the update in C1PO costs $O(R^{3})$ flops in the first iteration and $O(R^{2})$ flops in each subsequent
iteration. The computational complexity of the C2PO precoder reduces to $O(R^{2})$ in each iteration. It has been shown in \cite{Jedda2017Massive} that the number of arithmetic operations in each iteration of the MSM precoder is about ${\left( {2R + 1} \right)^2}\left( {2U + 4R} \right)$.
In the simulation, since the MSM precoder is based on linear programming, we use $linprog$ function in $Matlab$ platform to reproduce the MSM precoder. The proposed precoder only requires SVD of ${\tilde{\bf{H}}}^{\rm{T}}{\tilde{\bf{H}}}$ once in the first iteration
and the dominant computation load in each of the subsequent iterations is matrix-vector multiplication of complexity $O(R^2)$.

\begin{figure}[htbp]
\centering
\includegraphics[width=0.49\textwidth]{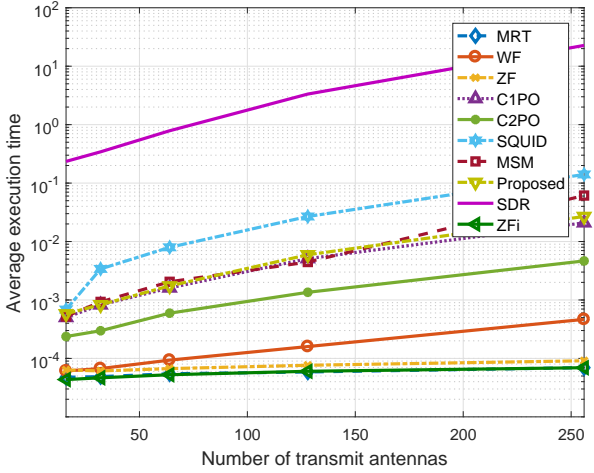}
\caption{Runtime comparison of the compared precoders versus the number of the BS antennas.}
\label{fig5}
\end{figure}

Fig. \ref{fig5} compares the average execution time (in seconds) of the algorithms versus different number of antennas.
The algorithms are performed on a desktop PC with an Intel Core I7-7700K CPU at 4.2 GHz with 32 GB RAM.
It can be seen that, the C2PO is the most efficient nonlinear precoder. The C1PO precoder, the MSM precoder and the proposed precoder have
comparable performance. The runtime of the SQUID and SDR precoders are respectively about
3.9 times and remarkably about 380 times that of the new precoder. On the other hand, the linear
quantized precoders, MRT, ZF and WF, have significantly lower complexity than the nonlinear
quantized precoders. However, as will be shown in the later experiments, the performance of the
linear quantized precoders are significantly worse than that of the nonlinear quantized precoders.

\subsection{Uncoded BER Performance}

We compare the performance of the new precoder and other involved precoders in the view of
uncoder BER with QPSK signaling.

\begin{figure}[htbp]
\centering
\includegraphics[width=0.49\textwidth]{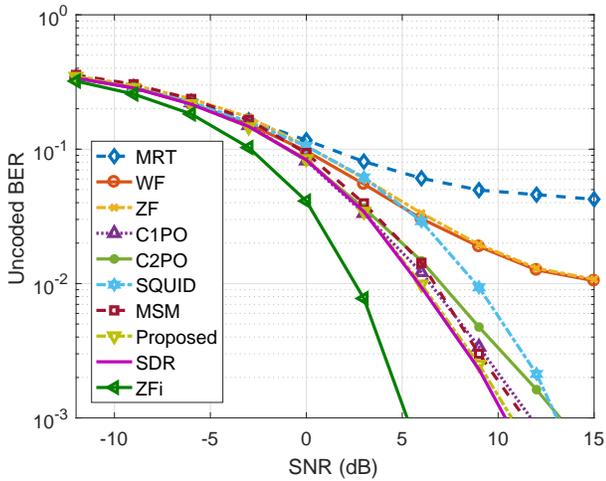}
\caption{Uncoded BER of the compared precoders in the case of a small size MU-MIMO system (16 BS antennas and 4 UTs) with QPSK signaling.}
\label{fig1a}
\end{figure}

Fig. \ref{fig1a} shows the performance of the precoders in the case of a
small size MU-MIMO system equipped with 16 BS antennas and serving 4 UTs.
It can be seen that, the performance of all involved nonlinear quantized precoders
surpasses that of the linear quantized precoders, MRT, ZF and WF, in most cases.
The benefit of using a nonlinear quantized precoder is especially conspicuous from
moderate to relatively high SNR, as the advantage of the nonlinear quantized precoders
over the linear quantized precoders increases as the SNR increases.

\begin{figure}[htbp]
\centering
\includegraphics[width=0.49\textwidth]{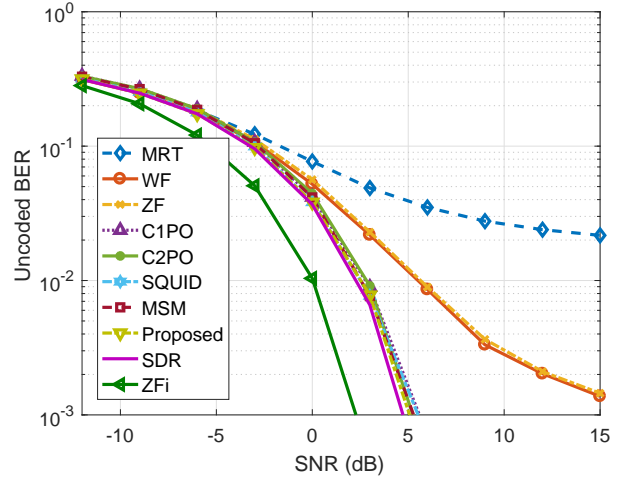}
\caption{Uncoded BER of the compared precoders in the case of a large size MU-MIMO system (128 BS antennas and 20 UTs) with QPSK signaling.}
\label{fig1b}
\end{figure}

Fig. \ref{fig1b} presents the result in the case of a massive MU-MIMO with 128 BS antennas and 20
UTs. It can be seen from Fig. \ref{fig1b} that, the linear quantized precoders tend to be saturated as the
SNR increases. Specially, the BERs do not decease distinctly as the SNR increases when the SNR
is above a certain threshold. In comparison, the nonlinear quantized precoders have significantly
better performance from moderate to relatively high SNR. Accordingly, the performance loss
due to 1-bit DAC cannot be compensated only by increasing the number of transmit antennas
but also need advanced signal processing techniques (e.g., nonlinear precoding).

From Fig. \ref{fig1a} and Fig. \ref{fig1b}, the performance of each precoder improves as the number
of the BS antennas increases. The SDR precoder and the proposed precoder
generally perform comparably in most cases.
In the case of a small size MU-MIMO system with 16 antennas,
the SDR precoder and the proposed precoder significantly outperform
other nonlinear precoder, e.g., when SNR $>$ 0 dB.
But in the case of a large size MU-MIMO system with 128 antennas,
the advantage of the SDR precoder and the proposed precoder over
the other nonlinear precoders is no longer so prominent and becomes marginal.

In short, the results in Fig. \ref{fig1a} and Fig. \ref{fig1b} imply that
the disparities between the ideal MU-MIMO system and the system with 1-bit DACs can be reduced by increasing the number of the BS antennas and by employing proper signal processing techniques. Besides, the results in Fig. \ref{fig5}, Fig. \ref{fig1a}, and Fig. \ref{fig1b} demonstrate that our method has comparable BER performance with SDR in both small- or large-scale MU-MIMO systems. The
C2PO is the most efficient, e.g., about 4 times faster than the new algorithm, but the new algorithm has
better BER performance, especially in the small-scale case.


\subsection{Effect of Different Modulation Modes}

\begin{figure}[htbp]
\centering
\includegraphics[width=0.49\textwidth]{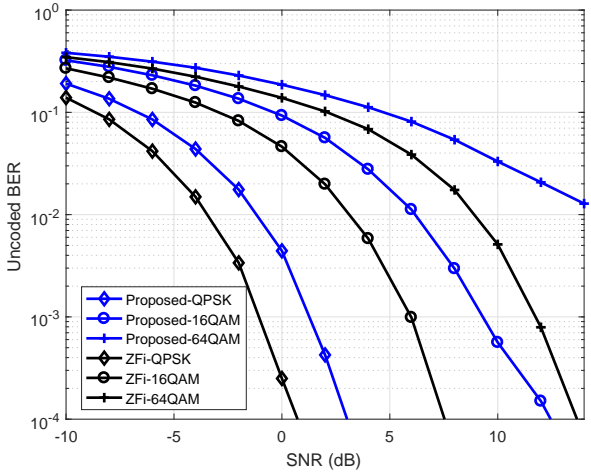}
\caption{Performance of the proposed precoder compared with the ideal linear precoder (ZFi) for different modulation modes.}
\label{fig2}
\end{figure}

This experiment investigates the proposed precoder for different modulation schemes,
QPSK, 16QAM and 64QAM. We consider a MU-MIMO system with 128 BS antennas and 10 UTs.
Fig. \ref{fig2} presents the performance of the proposed precoder compared with the
ideal ZFi precoder (with infinite-resolution DACs as benchmark) in the three modulation modes.
The results in Fig. \ref{fig2} indicate that the proposed precoder can provide reliable
transmission of QPSK signaling and well support the high modulation mode, e.g., 16QAM. But for the case of 64QAM,
reliable transmission is in need of more antennas and/or more sophisticated techniques
that we leave in our forthcoming work.

\subsection{Effect of the Channel Estimation Error}

We now investigate the effect of the channel estimation error.
Assume that the channel matrix can be model as

\begin{equation}
\hat{\bf{ H}} = \delta {\bf{H}} + \left( {1 - \delta } \right)\Delta {\bf{H}}
\end{equation}
where $\delta \in \left[ {0, 0.5} \right] $ accounts for the channel estimation factor.
$\delta=0$ implies the case of perfect CSI and $\delta > 0$ means imperfect CSI.
Two conditions with different distribution models are considered.
\textit{1) Gaussian model:} the entries of $\Delta {\bf{H}}$ are zero-mean
Gaussian distributed, ${\left[ {\Delta {\bf{H}}} \right]_{ij}} \sim \mathcal{CN}\left( {0,1} \right)$.
\textit{2) Non-Gaussian model:} the entries of $\Delta {\bf{H}}$ follow
a uniform distribution, ${\left[ {\Delta {\bf{H}}} \right]_{ij}} \sim \frac{1}{{\sqrt 3 }} \mathcal{UN}\left( { - 1,1} \right)$.

\begin{figure}[htbp]
\centering
\subfloat[Gaussian channel estimation error]{ \label{fig3a}
\includegraphics[width=0.49\textwidth]{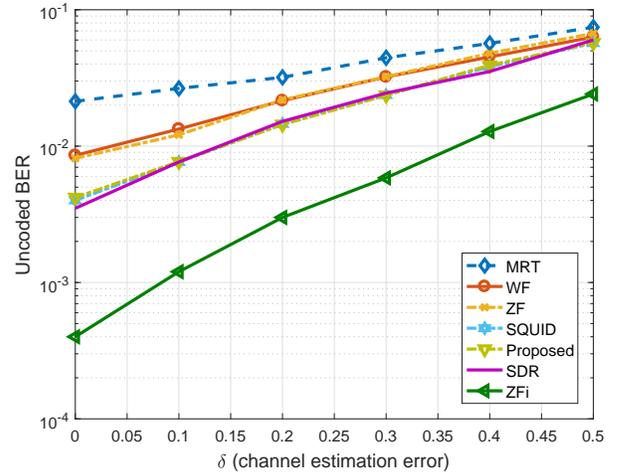}
}\\
\subfloat[Non-Gaussian channel estimation error]{ \label{fig3b}
\includegraphics[width=0.49\textwidth]{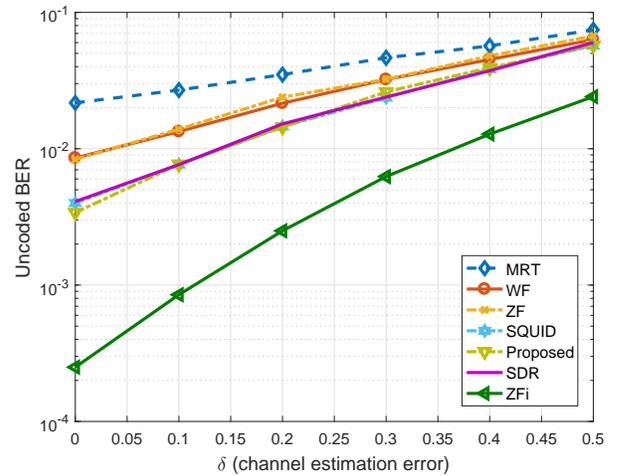}
}
\caption{Uncoded BER of the compared precoders for 1-bit MU-MIMO with imperfect CSI.}
\label{fig7}
\end{figure}

Fig. \ref{fig7} shows the performance of the compared precoders
versus the strength of the channel estimation error.
We consider a MU-MIMO system with 128 BS antennas and 16 UTs. The QPSK signaling is considered and the SNR is fixed at 0 dB.
The results indicate that, in both the Gaussian and non-Gaussian error conditions,
the nonlinear quantized precoders outperform the linear ones whenever $\delta > 0$.
It is demonstrated that, nonlinear quantized precoders can be adopted in
the case of imperfect CSI.

\subsection{Discussions}

For massive MIMO downlink with 1-bit DACs, we have proposed an efficient and robust nonlinear precoder, specifying no parameter
distribution of the channel matrix.  It is straight-forward to extend the results to the case of multiple bits DACs by leveraging the favorable relationship between the multi-bit DACs outputs and the single-bit ones (as shown in Section III-B of \cite{Chu2019R}). The codes we provided support massive MIMO systems with 1-3 bits DACs.

Furthermore, we have investigated the performance of the
proposed precoder under two types of channel estimation errors:
Gaussian error and non-Gaussian error. The Gaussian
assumption of the CSI is reasonable in the case of massive
MIMO with infinite-resolution ADCs \cite{Guo2011Estimation}.
In order to evaluate the robustness of the proposed algorithm in the conditions of non-Gaussian CSI error,
we considered a uniform distribution of the CSI error similar to \cite{Palomar2003Uniform}. For the case that massive MIMO systems are equipped with low-resolution ADCs and low-resolution DACs \cite{Kong2017Full, Kong2018Mul, Kong2018Non}, the distribution of channel estimation errors is different. In this paper, we restrict our work on the massive
MIMO downlink and leave out the consideration of low-resolution ADCs. In
future work, it is interesting to jointly consider both low-resolution ADCs and low-resolution DACs as \cite{Kong2018Mul, Kong2018Non}.

\section{Conclusions}
\label{Sec:f}

In this paper, we have investigated the nonlinear precoding problem for
MU-MIMO downlink systems using 1-bit DACs.
We proposed a highly-efficient algorithm based on the ADMM framework
to directly solve the nonconvex precoding problem. The proposed algorithm
is globally convergent under rather mild conditions.
A sufficient condition for its convergence has been derived.
In addition, various numerical studies have been conducted to verify
the effectiveness and efficiency of the new precoder. The superiority of the new precoder has
been demonstrated by various numerical studies. The results
showed that the proposed precoding algorithm can support the
MU-MIMO system with higher-modulation modes. Moreover,
it has comparable BER performance with SDR while being
much faster.

We end this section by posing an open problem arises in the MU-MIMO system with low-resolution DACs. Although present studies have shown that equipping MU-MIMO with 1-bit DACs is beneficial in terms of power consumption, the out-of-band (OOB) spectral regrowth \cite{Moll2015Out, Moll2018Out} may have an important impact on the performance of quantized MU-MIMO systems in practical setting. The OOB distortion can be reduced, to some extend, by increasing the number of the transmit antennas and the quantization level \cite{Jacobsson2018Nonlinear, Moll2018Out}. However, the OOB distortion issue has not been fully tackled
yet. It deserves further studies.

\section{Acknowledgements}
\label{Sec:g}

The authors would like to thank for technical experts: Tian
Pan, Guangyi Yang, Rui Gong and Yingzhe Li, all from Huawei Technologies for their fruitful discussions. We also
acknowledge the editor and the anonymous reviewers for their critical comments which help to improve the quality of the paper.

\begin{appendices}

\section{Derivation of \eqref{eqp9} }
\label{appdx}
The subproblem (16) is a form of
\begin{equation}
\label{eqap2}
{{\bf{u}}^{k + 1}} = \arg \mathop {{\rm{min}}}\limits_{\bf{u}} {\rm{ }}{\delta _\Omega }({\bf{u}}) + \frac{\lambda }{2}\left\| {{\boldsymbol{\omega }} - {\bf{u}}} \right\|_2^2.
\end{equation}
We consider a nondegenerate condition that $\left\| {\boldsymbol{\omega }} \right\|_2 > 0$ . The indicator function $\delta_{\Omega} \left( {{\bf{u}}} \right)$  enforces the solution to satisfy ${{\bf{u}}^{k + 1}} = a{\boldsymbol{\theta }}$ , with $a$ be a positive constant, ${\boldsymbol{\theta }} \in {\mathbb{R}^{2R}}$  with each element be 1 or -1. Then, problem \eqref{eqap2} can be equivalently rewritten as
\begin{equation}
\label{eqap1}
\begin{split}
{{\bf{u}}^{k + 1}} &= \arg \mathop {{\rm{min}}}\limits_{a,{\boldsymbol{\theta }}} {\rm{ }}\left\| {{\boldsymbol{\omega }} - a{\boldsymbol{\theta }}} \right\|_2^2\\
{\rm{subject \ to }} &~~\ a > 0, \ {\theta _i} \in \{  - 1, + 1\}
\end{split}
\end{equation}
for $i = 1,2, \cdots ,2R$.

Let  $I = {\rm{supp}}({\boldsymbol{\omega }})$ be the support of the nonzero elements in ${\boldsymbol{\omega }}$ , and ${I^c}$ denotes the complementary set of $I$  (which denotes the index set of the zero elements in  ${\boldsymbol{\omega }}$), we first prove \eqref{eqq20} by contradiction.

The objective function in \eqref{eqap1} can be equivalently expressed as
\[f(a,{\boldsymbol{\theta}}) = \sum\limits_{i \in I} {{{\left( {{w_i} - a{\theta _i}} \right)}^2}}  + \sum\limits_{i \in {I^c}} {{a^2}} .\]
Assume that there exists a solution $(a,{\bf{\tilde {\boldsymbol{\theta}} }})$ that ${\rm{sign}}({\tilde \theta _{{i_0}}}) \ne {\rm{sign}}({\tilde w_{{i_0}}})$  for some ${i_0} \in I$ , then, we can find another solution $(a,{\bf{\hat {\boldsymbol{\theta}} }})$  with  ${\rm{sign}}({\hat \theta _{{i_0}}}) = {\rm{sign}}({w_{{i_0}}})$ such that
\begin{align}
f(a,{{\tilde {\boldsymbol{\theta}} }}) &= \sum\limits_{i \in I} {{{\left( {{w_i} - a{{\tilde \theta }_i}} \right)}^2}}  + \sum\limits_{i \in {I^c}} {{a^2}} \notag \\
 &> \sum\limits_{i \in I} {{{\left( {{w_i} - a{{\hat \theta }_i}} \right)}^2}}  + \sum\limits_{i \in {I^c}} {{a^2}}  = f(a,{\bf{\hat {\boldsymbol{\theta}} }}) \notag
\end{align}
since ${({w_{{i_0}}} - a{\tilde \theta _{{i_0}}})^2} > {({w_{{i_0}}} - a{\hat \theta _{{i_0}}})^2}$  with $a > 0$. This contradicts to that   $(a,{{\tilde {\boldsymbol{\theta}} }})$ is a minimizer. Thus,  ${\boldsymbol{\theta}}$ satisfies \eqref{eqq20}.
With ${\boldsymbol{\theta}}$, the solution of  $a$ is explicitly given by
\[a = {\left( {{{\boldsymbol{\theta }}^{\rm{T}}}{\boldsymbol{\theta }}} \right)^{ - 1}}{{\boldsymbol{\theta }}^{\rm{T}}}{\boldsymbol{\omega}} = \frac{{{{\boldsymbol{\theta }}^{\rm{T}}}{\boldsymbol{\omega}}}}{{2R}}.\]
Further, since the elements in ${\boldsymbol{\theta}}$  satisfy \eqref{eqq20}, we finally obtain \eqref{eqp9}.

\section{Proof of Lemma I}
\label{append1}
First, the minimizer ${{\bf{u}}^{k + 1}}$ of \eqref{eq162} satisfies
\begin{equation}
\label{ap1}
\mathcal{L}({\tilde{\bf{ v}}^k},{{\bf{u}}^{k + 1}},{{\bf{w}}^k}) \le \mathcal{L}({\tilde{\bf{ v}}^k},{{\bf{u}}^k},{{\bf{w}}^k}).
\end{equation}
Let ${\rm{ei}}{{\rm{g}}_{\min }}({\tilde{\bf{H}}}^{\rm{T}}{\tilde{\bf{H}}})$  denote the minimal eigenvalue of ${\tilde{\bf{H}}}^{\rm{T}}{\tilde{\bf{H}}}$.
Since ${\tilde{\bf{H}}}^{\rm{T}}{\tilde{\bf{H}}}$ is a rank deficient matrix in the considered 1-bit precoding problem,
hence ${\rm{ei}}{{\rm{g}}_{\min }}({\tilde{\bf{H}}}^{\rm{T}}{\tilde{\bf{H}}})=0$ and it is easy to see that $\mathcal{L}(\tilde{\bf{ v}},{{\bf{u}}^{k + 1}},{{\bf{w}}^k})$ is $(2{\varphi _1} + 2c + \lambda )$--strongly convex with respect to $\tilde{\bf{ v}}$.  Thus, for any ${\tilde{\bf{ v}}^k}$, the minimizer  ${\tilde{\bf{ v}}^{k + 1}}$ of \eqref{eq161} satisfies
\begin{equation}
\label{ap2}
\begin{split}
&\mathcal{L}({{\tilde{\bf{ v}}}^{k + 1}},{{\bf{u}}^{k + 1}},{{\bf{w}}^k})\\
& \le \mathcal{L}({{\tilde{\bf{ v}}}^k},{{\bf{u}}^{k + 1}},{{\bf{w}}^k}) - \frac{{ 2c + \lambda }}{2}\left\| {{{\tilde{\bf{ v}}}^{k + 1}} - {{\tilde{\bf{ v}}}^k}} \right\|_2^2.
\end{split}
\end{equation}
Moreover, from the definition of $\mathcal{L}$ and with the use of \eqref{eq163}, we have
\begin{equation}
\label{ap3}
\begin{split}
&\mathcal{L}({{\tilde{\bf{ v}}}^{k + 1}},{{\bf{u}}^{k + 1}},{{\bf{w}}^{k + 1}}) - \mathcal{L}({{\tilde{\bf{ v}}}^{k + 1}},{{\bf{u}}^{k + 1}},{{\bf{w}}^k})\\
& = \frac{1}{\lambda }\left\| {{{\bf{w}}^{k + 1}} - {{\bf{w}}^k}} \right\|_2^2.
\end{split}
\end{equation}
Then, summing \eqref{ap1}, \eqref{ap2} and \eqref{ap3} yields
\begin{equation}
\label{ap4}
\begin{split}
&\mathcal{L}({{\tilde{\bf{ v}}}^{k + 1}},{{\bf{u}}^{k + 1}},{{\bf{w}}^{k + 1}}) - \mathcal{L}({{\tilde{\bf{ v}}}^k},{{\bf{u}}^k},{{\bf{w}}^k})\\
& \le \frac{1}{\lambda }\left\| {{{\bf{w}}^{k + 1}} - {{\bf{w}}^k}} \right\|_2^2 - \frac{{ 2c + \lambda }}{2}\left\| {{{\tilde {\bf{v}}}^{k + 1}} - {{\tilde{\bf{ v}}}^k}} \right\|_2^2.
\end{split}
\end{equation}
From \eqref{eq161}, the minimizer ${\tilde{\bf{ v}}^{k + 1}}$ given by each iteration satisfies
\begin{align}
\label{ap5}
2{{\tilde{\bf{H}}}^{\rm T}}({\tilde{\bf{H}}}{\tilde{\bf{ v}}^{k + 1}} - \tilde{\bf{ s}}) &+ 2c{\tilde{\bf{ v}}^{k + 1}} \notag \\
&~ + \lambda ({\tilde{\bf{ v}}^{k + 1}} - {{\bf{u}}^{k + 1}} - {{\bf{w}}^k}/\lambda ) = {\bf{0}}.
\end{align}

Substituting \eqref{eq163} into \eqref{ap5} we have
\begin{equation}
\label{ap6}
2{{\tilde{\bf{H}}}^{\rm T}}({\tilde{\bf{H}}}{\tilde{\bf{ v}}^{k + 1}} - \tilde{\bf{ s}}) + 2c{\tilde{\bf{ v}}^{k + 1}} - {{\bf{w}}^{k + 1}} = {\bf{0}}.
\end{equation}
Then, we have
\begin{equation}
\label{ap7}
\begin{split}
&\left\| {{{\bf{w}}^{k + 1}} - {{\bf{w}}^k}} \right\|_2^2\\
& = 4\left\| {({{\tilde{\bf{H}}}^{\rm T}}{\tilde{\bf{H}}} + c{\bf{I}})({{\tilde{\bf{ v}}}^{k + 1}} - {{\tilde{\bf{ v}}}^k})} \right\|_2^2\\
& \le 4{({\varphi} + c)^2}\left\| {{{\tilde{\bf{ v}}}^{k + 1}} - {{\tilde{\bf{ v}}}^k}} \right\|_2^2
\end{split}.
\end{equation}
Consequently, substituting \eqref{ap7} into \eqref{ap4} results in Lemma I.

\section{Proof of Lemma II}
\label{append2}

To prove Lemma II, we first show the sequence
$\{ ({\tilde{\bf{ v}}^k},{{\bf{u}}^k},{{\bf{w}}^k})\} $
generated via \eqref{eq161}--\eqref{eq163} is bounded if the condition
\eqref{equ:conv_cond} holds. From \eqref{ap6}, we have
\begin{equation}
\label{ap8}
\begin{split}
\left\| {{{\bf{w}}^k}} \right\|_2^2 &= 4\left\| {{{\tilde{\bf{H}}}^{\rm T}}({\tilde{\bf{H}}}{{\tilde{\bf{ v}}}^k} - \tilde{\bf{ s}}) + c{{\tilde{\bf{ v}}}^k}} \right\|_2^2\\
 &\le 4{\left( {{{\left\| {{{\tilde{\bf{H}}}^{\rm T}}({\tilde{\bf{H}}}{{\tilde{\bf{ v}}}^k} - \tilde{\bf{ s}})} \right\|}_2} + c{{\left\| {{{\tilde{\bf{ v}}}^k}} \right\|}_2}} \right)^2}\\
 &\le 8\left\| {{{\tilde{\bf{H}}}^{\rm T}}({\tilde{\bf{H}}}{{\tilde{\bf{ v}}}^k} - \tilde{\bf{ s}})} \right\|_2^2 + 8{c^2}\left\| {{{\tilde{\bf{ v}}}^k}} \right\|_2^2\\
 &\le 8{\varphi}\left\| {{\tilde{\bf{H}}}{{\tilde{\bf{ v}}}^k} - \tilde{\bf{ s}}} \right\|_2^2 + 8{c^2}\left\| {{{\tilde {\bf{v}}}^k}} \right\|_2^2
\end{split}
\end{equation}
where we have used the relation
\[{\varphi} = {\rm{ei}}{{\rm{g}}_{\max }}({{\tilde{\bf{H}}}^{\rm T}}{\tilde{\bf{H}}}) = {\rm{ei}}{{\rm{g}}_{\max }}({\tilde{\bf{H}}}{{\tilde{\bf{H}}}^{\rm T}}).\]
Note that, $\mathcal{L}({\tilde{\bf{ v}}^k},{{\bf{u}}^k},{{\bf{w}}^k})$
is bounded from below. When \eqref{equ:conv_cond} holds,
it follows from Lemma I that $\mathcal{L}({\tilde{\bf{ v}}^k},{{\bf{u}}^k},{{\bf{w}}^k})$
is non-increasing in the iteration and therefore convergent.
Then, from the definition of $\mathcal{L}$ and with the use of \eqref{ap8},
for any $k > 1$ we have
\begin{equation}\label{ap9}
\begin{split}
&{\mathcal{L}({{\tilde{\bf{ v}}}^1},{{\bf{u}}^1},{{\bf{w}}^1})}\\
& \ge \mathcal{L}({{\tilde{\bf{ v}}}^k},{{\bf{u}}^k},{{\bf{w}}^k})\\
& = \left\| {\tilde {\bf{s}} - \tilde{\bf{ H}}{{\tilde{\bf{ v}}}^k}} \right\|_2^2 + c\left\| {{{\tilde{\bf{ v}}}^k}} \right\|_2^2 + {\delta _\Omega }({{\bf{u}}^k})\\
&~~~ + \frac{\lambda }{2}\left\| {{{\tilde{\bf{ v}}}^k} - {{\bf{u}}^k} - \frac{{{{\bf{w}}^k}}}{\lambda }} \right\|_2^2 - \frac{{\left\| {{{\bf{w}}^k}} \right\|_2^2}}{{2\lambda }}\\
& \ge {c_1}\left\| {\tilde {\bf{s}} - \tilde{\bf{ H}}{{\tilde {\bf{v}}}^k}} \right\|_2^2 + {c_2}\left\| {{{\tilde{\bf{ v}}}^k}} \right\|_2^2 + {\delta _\Omega }({{\bf{u}}^k})\\
&~~~ + \frac{\lambda }{2}\left\| {{{\tilde{\bf{ v}}}^k} - {{\bf{u}}^k} - \frac{{{{\bf{w}}^k}}}{\lambda }} \right\|_2^2
\end{split}
\end{equation}
with ${c_1} = 1 - 8{\varphi}/\lambda $ and ${c_2} = c - 8{c^2}/\lambda $.
It is easy to see that, when \eqref{equ:conv_cond} holds, ${c_1} > 0$  and ${c_2} > 0$,
then it follows from \eqref{ap9} that the sequence
$\{ ({\tilde{\bf{ v}}^k},{{\bf{u}}^k},{{\bf{w}}^k})\} $ is bounded.
Since the sequence ${{\bf{z}}^k}: = ({\tilde{\bf{ v}}^k},{{\bf{u}}^k},{{\bf{w}}^k})$
is bounded, there exists a convergent subsequence ${{\bf{z}}^{{k_j}}}$
that converges to a cluster point ${{\bf{z}}^{\rm{*}}}$.
Meanwhile, from Lemma I, $\mathcal{L}({{\bf{z}}^k})$ is nonincreasing
and convergent when \eqref{equ:conv_cond} holds, and
$\mathcal{L}({{\bf{z}}^k}) \ge \mathcal{L}({{\bf{z}}^*})$ for any $k$.
Thus, using the result in Lemma 1 we have
\begin{equation}
\begin{split}
\infty  &> \mathcal{L}({{\bf{z}}^1}) - \mathcal{L}({{\bf{z}}^*}) \\
     &\ge \mathcal{L}({{\bf{z}}^1}) - \tilde {\mathcal{L}}({{\bf{z}}^{K + 1}}) \\
     &= \sum\limits_{k = 1}^K {\left[ {\mathcal{L}({{\bf{z}}^k}) - \mathcal{L}({{\bf{z}}^{k + 1}})} \right]}\\
     &\ge \left[ {\frac{{ 2c + \lambda }}{2} - \frac{{4{{({\varphi} + c)}^2}}}{\lambda }} \right]\sum\limits_{k = 1}^K {\left\| {{{\tilde{\bf{ v}}}^{k + 1}} - {{\tilde{\bf{ v}}}^k}} \right\|_2^2} .\notag
\end{split}
\end{equation}
Let  $K \to \infty $, when \eqref{equ:conv_cond} holds, we have
\begin{equation}
\frac{{2c + \lambda }}{2} - \frac{{4{{({\varphi} + c)}^2}}}{\lambda } > 0\notag
\end{equation}
and it follows that
\begin{equation}\label{ap10}
\sum\limits_{k = 1}^\infty  {\left\| {{{\tilde {\bf{v}}}^{k + 1}} - {{\tilde {\bf{v}}}^k}} \right\|_2^2}  < \infty
\end{equation}
which together with \eqref{ap7} and \eqref{eq163} implies
\begin{equation}
\label{ap11}
\sum\limits_{k = 1}^\infty  {\left\| {{{\bf{w}}^{k + 1}} - {{\bf{w}}^k}} \right\|_2^2}  < \infty
\end{equation}
and
\begin{equation}
\label{ap12}
\sum\limits_{k = 1}^\infty  {\left\| {{{\bf{u}}^{k + 1}} - {{\bf{u}}^k}} \right\|_2^2}  < \infty.
\end{equation}
Then, from \eqref{ap10}, \eqref{ap11}, and \eqref{ap12} we have
\[\mathop {\lim }\limits_{k \to \infty } \left\| {{{\bf{z}}^{k + 1}} - {{\bf{z}}^k}} \right\|_2^2 = 0.\]

Next, we show that any cluster point of the sequence ${\rm{\{ }}{{\bf{z}}^k}{\rm{\} }}$
is a stationary point of $\mathcal{L}$. The sequence generated via the steps \eqref{eq161}--\eqref{eq163} satisfies
\begin{equation}
\label{ap13}
\left\{ \begin{array}{l}
{\bf{0}} \in \partial {\delta _\Omega }({{\bf{u}}^{k + 1}}) + {{\bf{w}}^{k + 1}} + \lambda ({{\tilde{\bf{ v}}}^{k + 1}} - {{\tilde{\bf{ v}}}^k})\\
{\bf{0}} = 2({{\tilde{\bf{H}}}^{\rm T}}{\tilde{\bf{H}}} + c{\bf{I}}){{\tilde{\bf{ v}}}^{k + 1}} - 2{{\tilde{\bf{H}}}^{\rm T}}\tilde{\bf{ s}} - {{\bf{w}}^{k + 1}}\\
{{\bf{w}}^{k + 1}} = {{\bf{w}}^k} - \lambda ({{\tilde{\bf{ v}}}^{k + 1}} - {{\bf{u}}^{k + 1}})
\end{array} \right..
\end{equation}
Let $\{ {{\bf{z}}^{{k_j}}}\}$  denote a convergent subsequence of $\{ {{\bf{z}}^k}\} $,
since $\mathop {\lim }\limits_{k \to \infty } \left\| {{{\bf{z}}^{k + 1}} - {{\bf{z}}^k}} \right\|_2^2 = 0$,
the two sequences $\{ {{\bf{z}}^{{k_j}}}\} $  and  $\{ {{\bf{z}}^{{k_j} + 1}}\} $ have the same limit point,
denoted by ${{\bf{z}}^*}: = ({\tilde{\bf{ v}}^*},{{\bf{u}}^*},{{\bf{w}}^*})$.
When the condition \eqref{equ:conv_cond} is satisfied, $\mathcal{L}({{\bf{z}}^k})$ is convergent,
thus ${\delta _\Omega }({{\bf{u}}^{k + 1}})$ is also convergent.
Then, passing to the limit in \eqref{ap13} along the subsequence $\{ {{\bf{z}}^{{k_j}}}\} $ yields
\[\left\{ \begin{array}{l}
 - {{\bf{w}}^*} \in \partial {\delta _\Omega }({{\bf{u}}^*})\\
{{\bf{w}}^*} = 2({{\tilde{\bf{H}}}^{\rm T}}{\tilde{\bf{H}}} + c{\bf{I}}){{\tilde{\bf{ v}}}^*} - 2{{\tilde{\bf{H}}}^{\rm T}}\tilde{\bf{ s}}\\
{{\tilde{\bf{ v}}}^*} = {{\bf{u}}^*}
\end{array} \right..\]
Consequently, such a limit point ${{\bf{z}}^*}$  is a
stationary point of the Lagrangian function $\mathcal{L}$.

\section{Proof of Lemma III}
\label{append3}

From the definition of the Lagrangian function   in \eqref{eq162}, we have
\[{\partial _{\bf{u}}}\mathcal{L}({{\bf{z}}^{k{\rm{ + 1}}}}) = \partial {\delta _\Omega }({{\bf{u}}^{k + 1}}) - ({{\bf{w}}^{k + 1}} - {{\bf{w}}^k}) + {{\bf{w}}^{k + 1}}\]
which together with the first relation in \eqref{ap13} yields
\[\lambda ({\tilde{\bf{ v}}^{k + 1}} - {\tilde{\bf{ v}}^k}) - ({{\bf{w}}^{k + 1}} - {{\bf{w}}^k}) \in {\partial _{\bf{u}}}\mathcal{L}({{\bf{z}}^{k{\rm{ + 1}}}}).\]
Similarly, we have
\[{\partial _{\tilde{\bf{ v}}}}\mathcal{L}({{\bf{z}}^{k{\rm{ + 1}}}}) = 2({{\tilde{\bf{H}}}^{\rm T}}{\tilde{\bf{H}}} + c{\bf{I}}){\tilde{\bf{ v}}^{k + 1}} - 2{{\tilde{\bf{H}}}^{\rm T}}\tilde{\bf{ s}} - {{\bf{w}}^k} \]
which together with the second relation in \eqref{ap13} implies
\[{\partial _{\tilde{\bf{ v}}}}\mathcal{L}({{\bf{z}}^{k{\rm{ + 1}}}}) = {{\bf{w}}^{k + 1}} - {{\bf{w}}^k}.\]
Meanwhile, we have
\[{\partial _{\bf{w}}}\mathcal{L}({{\bf{z}}^{k{\rm{ + 1}}}}) = ({{\bf{w}}^{k + 1}} - {{\bf{w}}^k})/\lambda. \]
Thus, there exists a constant ${c_2} > 0$  such that
\begin{equation}
{\rm{dist}}(0,\partial \mathcal{L}({{\bf{z}}^{k{\rm{ + 1}}}})) \le
{c_2}\left( {\left\| {{{\tilde{\bf{ v}}}^{k + 1}} - {{\tilde{\bf{ v}}}^k}} \right\|_2^2 + \left\| {{{\bf{w}}^{k + 1}} - {{\bf{w}}^k}} \right\|_2^2} \right) \notag
\end{equation}
which together with \eqref{ap7} results in Lemma III.

\section{Proof of Theorem I}
\label{append4}

With the results in Lemma II and Lemma III, the rest proof of Theorem I is
to prove that the sequence ${\rm{\{ }}{{\bf{z}}^k}{\rm{\} }}$ generated by
the ADMM algorithm is a $Cauchy \ sequence$ and has finite length

\begin{equation}
\sum\limits_{k = 0}^\infty  {{{\left\| {{{\bf{z}}^{k + 1}} - {{\bf{z}}^k}} \right\|}_2}}  < \infty.\notag
\end{equation}
Then, the global convergence of the sequence ${\rm{\{ }}{{\bf{z}}^k}{\rm{\} }}$
to a stationary point of $\mathcal{L}$ is proved. The derivation of this property
relies on the Kurdyka-Lojasiewicz (KL) property of $\mathcal{L}$. $\mathcal{L}$ is
a KL function since it is a composite of an indicator function and analytical function.
The rest derivation is omitted here for succinctness since it follows similarly to
that in \cite{li2015global} with only some minor changes.

\end{appendices}

\bibliographystyle{IEEEtran}
\normalem
\bibliography{icc_leo}

\begin{IEEEbiography}[{\includegraphics[width=1in,height=2.5in,clip,keepaspectratio]{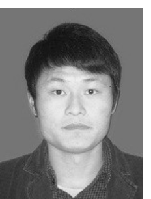}}]
{Lei Chu}
has been pursuing the Ph.D degree at Shanghai Jiaotong University, since 2015. He is a research assistant in the Big Data Engineering Technology and Research Center, Shanghai. He has published two book chapters, over 20 papers in refereed journals/conferences. He serves a reviewer for many journal papers.

His research interests are in the theoretical and algorithmic studies in random matrix theory, non-convex optimization, deep learning, as well as their applications in wireless communications, bioengineering, and smart grid.
\end{IEEEbiography}

\begin{IEEEbiography}[{\includegraphics[width=1in,height=2.5in,clip,keepaspectratio]{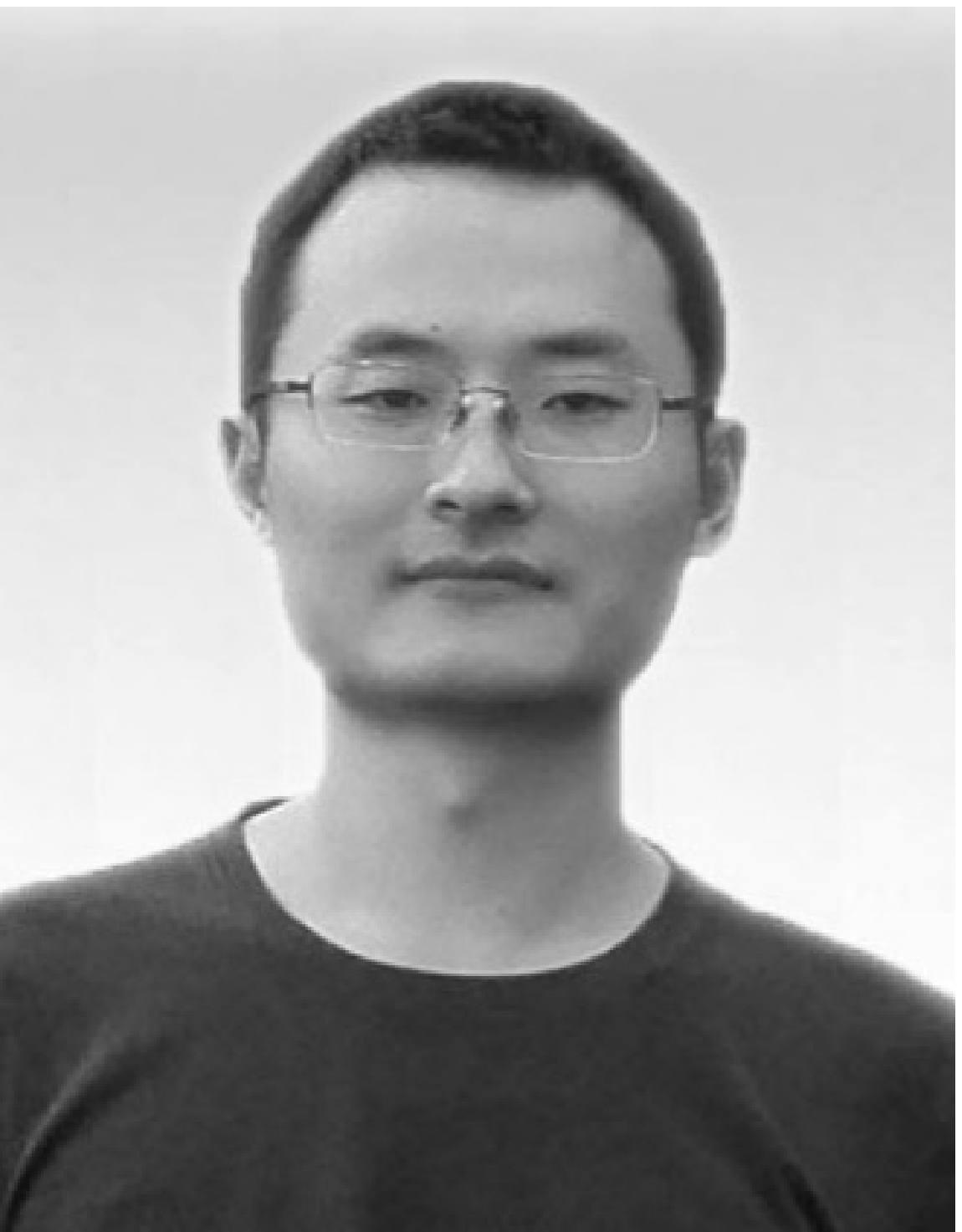}}]
{Fei Wen}
received the B.S. degree from the University of Electronic Science and Technology of China (UESTC) in 2006, and the Ph.D. degree in communications and information engineering from UESTC in 2013. Now he is a research associate professor in the Department of Electronic Engineering of Shanghai Jiao Tong University.

His main research interests are nonconvex optimization, large-scale numerical optimization, sparse and statistical signal processing.
\end{IEEEbiography}

\begin{IEEEbiography}[{\includegraphics[width=1in,height=2in,clip,keepaspectratio]{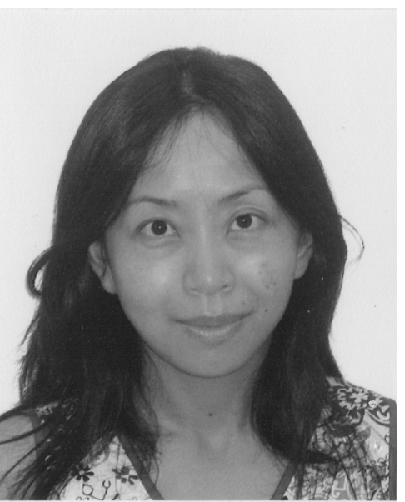}}]
{Lily Li}
(S'11) received the B.S. degree in computer science from University of Electronic and Science Technology of China, Chengdu, China,  in 1992 and the M.S. degree in computer science in New York University, New York, USA, in 2000 and the M.S. degree in mathematics in Tennessee Technological University, Cookeville, TN, USA in 2010. She is current working toward the Ph.D. degree in Electrical and computer engineering in Tennessee Technological University.

Her research interests include statistics, machine learning, big data and wireless communications
\end{IEEEbiography}

\begin{IEEEbiography}[{\includegraphics[width=1in,height=2.5in,clip,keepaspectratio] {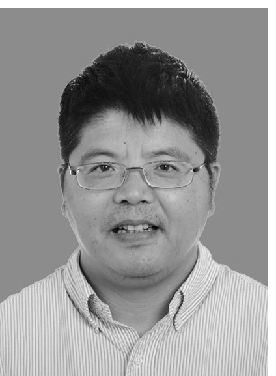}}]
{Robert Qiu}
(IEEE S'93-M'96-SM'01-FM¡¯14) received the Ph.D. degree in electrical engineering from New York University (former Polytechnic University, Brooklyn, NY). He  is currently a Professor in the Department of Electrical and Computer Engineering, Center for Manufacturing Research, Tennessee Technological University, Cookeville, Tennessee, where he started as an Associate Professor in 2003 before he became a Professor in 2008.  He has also been with the Department of Electrical Engineering, Research Center for Big Data Engineering and Technologies, State Energy Smart Grid R$\&$D Center, Shanghai Jiaotong University since 2015.

He was Founder-CEO and President of Wiscom Technologies, Inc., manufacturing and marketing WCDMA chipsets. Wiscom was sold to Intel in 2003. Prior to Wiscom, he worked for GTE Labs, Inc. (now Verizon), Waltham, MA, and Bell Labs, Lucent, Whippany, NJ. He has worked in wireless communications and network, machine learning, Smart Grid, digital signal processing, EM scattering, composite absorbing materials, RF microelectronics, UWB, underwater acoustics, and fiber optics. He holds over 6 patents and authored over 70 journal papers/book chapters and 90 conference papers. He has 15 contributions to 3GPP and IEEE standards bodies. In 1998 he developed the first three courses on 3G for Bell Labs researchers. He served as an adjunct professor in Polytechnic University, Brooklyn, New York. Dr. Qiu serves as Associate Editor, IEEE TRANSACTIONS ON VEHICULAR TECHNOLOGY and other international journals. He is a co-author of Cognitive Radio Communication and Networking: Principles and Practice (John Wiley), 2012 and Cognitive Networked Sensing: A Big Data Way (Springer), 2013, and the author of Big Data and Smart Grid (John Wiley), 2015. He is a Guest Book Editor for Ultra-Wideband (UWB) Wireless Communications (New York: Wiley, 2005), and three special issues on UWB including the IEEE JOURNAL ON SELECTED AREAS IN COMMUNICATIONS, IEEE TRANSACTIONS ON VEHICULAR TECHNOLOLOGY and IEEE TRANSACTION ON SMART GRID. He serves as a Member of TPC for GLOBECOM, ICC, WCNC, MILCOM, ICUWB, etc. In addition, he served on the advisory board of the New Jersey Center for Wireless Telecommunications (NJCWT).

He was named a Fellow of the Institute of Electrical and Electronics Engineers (IEEE) in 2015 for his contributions to ultra-wideband wireless communications. His current interests are in wireless communication and networking, random matrix theory based theoretical analysis for deep learning,  and the Smart Grid technologies.

\end{IEEEbiography}

\end{document}